\documentclass[prl,twocolumn,showpacs,nofootinbib,aps,10pt,superscriptaddress,amsmath,amssymb]{revtex4-1}

\usepackage{graphicx}% Include figure files
\usepackage{dcolumn}% Align table columns on decimal point
\usepackage{bm}% bold math
\usepackage{epsfig}
\usepackage{epstopdf}
\usepackage{mathtools}
\usepackage{natbib}
\usepackage{captcont}
\usepackage{booktabs}
\usepackage{threeparttable}
\usepackage{slashed}
\usepackage{extarrows}

\begin{document}

\title{Width effects in resonant three-body decays: $B$ decay as an example}

\author{Hai-Yang Cheng}
\email[e-mail: ]{phcheng@phys.sinica.edu.tw}
\affiliation{Institute of Physics, Academia Sinica, Taipei, Taiwan 11529, R.O.C.}

\author{Cheng-Wei Chiang}
\email[e-mail: ]{chengwei@phys.ntu.edu.tw}
\affiliation{Department of Physics and Center for Theoretical Physics, National Taiwan University,
Taipei, Taiwan 10617, R.O.C.}
\affiliation{Physics Division, National Center for Theoretical Sciences,
Taipei, Taiwan 10617, R.O.C.}

\author{Chun-Khiang Chua}
\email[e-mail: ]{ckchua@cycu.edu.tw}
\affiliation{Department of Physics and Center for High Energy Physics, Chung-Yuan Christian University, Chung-Li, Taiwan 32023, R.O.C.}

\begin{abstract}

For three-body hadron decays mediated by intermediate resonances with large widths, we show how to properly extract the quasi-two-body decay rates for a meaningful comparison with the corresponding theoretical estimates, using several $B$ decays as explicit examples.  We compute the correction factor from finite width effects in the QCD factorization approach, and make a comparison with that using the experimental parameterization in which the momentum dependence in the weak dynamics is absent.  Although the difference is generally less than $10\%$ for tensor and vector resonances, it can be as large as $(25-40)\%$ for scalar resonances.  Our finding can in fact be applied to general quasi-two-body decays.

\end{abstract}
\maketitle

{\it Introduction ---}
Hadronic decays of heavy mesons have been of great interest to physicists  because they provide an ideal environment to test our understanding of heavy flavor symmetry, effective weak interactions, and strong interactions at low energies.  Among such decays, quasi-two-body decays form an important class in the study of bottom and charm meson physics.  They typically involve an unstable particle in the final state that further decays to two hadrons, possibly through strong interactions, and result in a three-body final state.  In the resonant region around the pole mass of the intermediate particle, the contribution of quasi-two-body decay overwhelms the non-resonant channels.

Take a $B$ meson decay $B\to RP_3\to P_1P_2P_3$ as an example, where $R$ and $P_3$ are respectively an intermediate resonant state and a hadron and $R$ further decays to two hadrons $P_{1,2}$.  It is a common practice to apply the factorization relation, also known as the narrow width approximation (NWA), to factorize the process as two sequential two-body decays:
\begin{align}
\label{eq:fact}
\begin{split}
&
{\cal B}(B\to RP_3\to P_1P_2P_3)
\\
&\qquad\qquad
={\cal B}(B\to RP_3){\cal B}(R\to P_1P_2)
~,
\end{split}
\end{align}
thereby extracting the branching fraction ${\cal B}(B\to RP_3)$ from the other two measured branching fractions, which is then compared with theoretical predictions.  However, such a treatment is valid only in the limit of $\Gamma_R\to 0$.  In other words, one should have instead ${\cal B}(B\to RP_3\to P_1P_2P_3)_{\Gamma_R\to 0}$ on the left-hand side of Eq.~\eqref{eq:fact}, assuming that both ${\cal B}(B\to RP_3)$ and ${\cal B}(R\to P_1P_2)$ are not affected by the NWA. In particular, all the decay channels of $R$ vanish in the zero width limit in such a way that ${\cal B}(R\to P_1P_2)$ remains intact.  When $R$ assumes a finite width, Eq.~(\ref{eq:fact}) no longer holds.  Besides, theoretical predictions of $\Gamma(B\to RP_3)$ are normally calculated under the assumption that both final-state particles are stable ({\it i.e.}, $\Gamma_R , \Gamma_{P_3} \to 0$).  Therefore, the question is for $R$ with a sufficiently large width, how one should extract ${\cal B}(B\to RP_1)$ from the experimental measurement of ${\cal B}(B\to RP_3\to P_1P_2P_3)$ and make a meaningful comparison with its theoretical predictions.

We propose that one should use
\begin{align}
\label{eq:BRofRP}
{\cal B}(B\to RP_3)
= \eta_{_R} \frac{{\cal B}(B\to RP_3\to P_1P_2P_3)_{\rm exp}}{{\cal B}(R\to P_1P_2)}
~,
\end{align}
where ${\cal B}(B\to RP_3\to P_1P_2P_3)_{\rm exp}$ is the measured branching fraction around the resonance region and the correction factor is defined by
\begin{align}
\label{eq:eta}
\eta_{_R}\equiv
\frac{ {\cal B}(B\to RP_3\to P_1P_2P_3)_{\Gamma_R\to 0}}{{\cal B}(B\to RP_3\to P_1P_2P_3)}
~.
\end{align}
Here ${\cal B}(B\to RP_3)$ on the left-hand side of Eq.~\eqref{eq:BRofRP} is the branching fraction assuming that both $R$ and $P_3$ are stable and thus have zero decay width.  Therefore, it is suitable for a comparison with theoretical calculations.

Conventionally, $\eta_R$ is set as unity in Eq.~\eqref{eq:BRofRP} to extract ${\cal B}(B\to RP_3)$ in the literature ({\it e.g.}, the Particle Data Group~\cite{PDG}).  This is inappropriate because $\eta_R \neq 1$ in general.  The parameter $\eta_R$ is calculable; it depends on not only the resonant state $R$ but also the decay process.  It is also noted that $\eta_R$ and the corresponding quantity for the {\it CP}-conjugated process are generally different as a result of the interference between decay amplitudes with different {\it CP}-violating phases, though the differences for the modes we have studied are quite small.

While we focus on some three-body $B$ meson decays in this Letter to elucidate our point and explain the cause, our finding generally applies to all quasi-two-body decays.  To compute $\eta_R$, we employ the QCD factorization (QCDF)~\cite{BBNS} to model both numerator and denominator on the right-hand side of Eq.~\eqref{eq:eta}.  The results are compared with those obtained using the experimental parameterization (EXPP).  A more detailed numerical study has been presented elsewhere~\cite{CCC}.

{\it General framework ---}
Consider the $B\to RP_3\to P_1P_2P_3$ decay, where $R$ is a resonance of spin $J$.  The associated decay amplitude has the form
\begin{align}
\label{eq: A generic}
\!\!\!\!
A(m_{12}, m_{23})
=
{\cal M}(m_{12}, m_{23}) R_J(m_{12}) {\cal T}_J(m_{12},m_{23})
~,
\end{align}
where $m^2_{ij}\equiv (p_i+p_j)^2$, ${\cal M}(m_{12}, m_{12})$ is a function containing information of both $B\to R(m_{12})P_3$ and $R(m_{12})\to P_1P_2$ decays, $R_J$ describes the line shape of the resonance and ${\cal T}_J$ encodes the angular dependence.  For example, a common choice of $R_J$ is the relativistic Breit-Wigner line shape
\vspace{-1mm}
\begin{align}
R_J^{\rm BW}(m_{12})=\frac{1}{(m^2_{12}-m_R^2)+i m_R \Gamma_R(m_{12})}
~.
\label{eq: RJ BW}
\end{align}
At the resonance ($m_{12} = m_R$), the amplitude can be factorized into the form
\begin{align}
\begin{split}
&
i\sqrt{\pi m_R \Gamma_R} \, A(m_R,m_{23})
\\
&=\frac{\sum_\lambda {\cal M}_\lambda[B\to R(m_R) P_3] {\cal M}_\lambda[R(m_R)\to P_1 P_2]}
{\sqrt{m_R \Gamma_R/\pi}} ~,
\end{split}
\label{eq: A at mR generic}
\end{align}
where $\lambda$ is the helicity of $R$.
The above form is expectable because there is a propagator of $R$ in the amplitude $A(m_{12}, m_{23})$, and the denominator of the propagator reduces to $i m_R \Gamma_R$ on the mass shell of $m_{12}$ while the numerator reduces to a polarization sum of the polarization vectors, producing the above structure after contracting with the rest of the amplitude.  The angular distribution term ${\cal T}_J$ in Eq. (\ref{eq: A generic}) at the resonance is governed by the Legendre polynomial $P_J(\cos\theta)$, where $\theta$ is the angle between $\vec{p}_1$ and $\vec{p}_3$ measured in the rest frame of $R$ (see also \cite{Asner:2003gh}), as a result of the polarization sum and the addition theorem of spherical harmonics.

Using the standard formulas~\cite{PDG}, the three-body differential decay rate at the resonance after the $dm_{23}^2$ phase space integration can be recast to give
\vspace{-1mm}
\begin{align}
\begin{split}
&
\pi m_R \Gamma_R \frac{d\Gamma(m^2_R)}{d m^2_{12}}
\\
&=\Gamma(B\to R P_3){\cal B}(R\to P_1 P_2)
\\
&=
\frac{\pi m_R \Gamma_R}{(2\pi)^3}\frac{1}{32 m_B^3}
\int |A(m_R,m_{23})|^2 dm_{23}^2
~,
\end{split}
\label{eq: RP3 RPP generic}
\end{align}
where we have made use of the fact that ${\cal B}(R\to P_1 P_2)$ is independent of the helicity (or spin) and $\Gamma_R$ of $R$.  Note that the ostensible $\Gamma_R$ dependence on the right-handed side of Eq.~\eqref{eq: RP3 RPP generic} would be canceled out by a corresponding factor coming from the phase space integral.
With the definition of normalized differential rate (NDR)
\begin{align}
\frac{d\tilde\Gamma (m_{12}^2)}{dm^2_{12}}
\equiv
{\frac{d\Gamma (m_{12}^2)}{dm^2_{12}}}
\bigg/
{\int \frac{d\Gamma (m_{12}^2)}{dm^2_{12}} dm_{12}^2}
~,
\label{eq: normalized dGamma}
\end{align}
we then have for the correction factor that
\begin{align}
\begin{split}
\eta_R
&=
\frac{\pi m_R \Gamma_R
\int
 |A(m_R, m_{23})|^2 dm_{23}^2}{\int  |A(m_{12}, m_{23})|^2 dm_{12}^2 \, dm_{23}^2}
 \\
&=
\pi m_R \Gamma_R\, \frac{d\tilde\Gamma (m_R^2)}{dm^2_{12}}
= \frac{\pi\Gamma_R}{2} \frac{d\tilde\Gamma (m_R)}{dm_{12}}
~.
\end{split}
\label{eq: eta A}
\end{align}
This equation shows that $\eta_R$ is determined by the value of the NDR at the resonance.  It should be noted that as the NDR is always positive and normalized to unity after integration, its value at  $m_{12} = m_R$ is anticorrelated to that elsewhere.  Hence, it is the shape of the NDR that matters in the determination of $\eta_R$.
With the help of the following identity
\begin{align}
\lim_{\Gamma_R\to 0} \frac{m_R \Gamma_R/\pi}{(m^2_{12}-m_R^2)^2+m^2_R\Gamma^2_R}
=\delta(m^2_{12}-m^2_R)
~,
\label{eq: identity}
\end{align}
one can readily verify that $\eta_R$ given by \eqref{eq: eta A}
approaches to unity in the NWA.  It has also been checked that the same would be obtained if one chose to use the Gounaris-Sakurai line shape~\cite{Gounaris:1968mw} instead.

The following parameterization is widely used in the experimental studies of resonant three-body $B$ decays
~\cite{Asner:2003gh}
\begin{align} \label{eq:cF}
M(m_{12}, m_{23})
=c X_J(p_3)\times X_J(p_1),
\end{align}
with $X_J$ being the Blatt-Weisskopf barrier form factor
and $c$ a complex coefficient to be determined experimentally.
It is straightforward to obtain $\eta^{\rm EXPP}_R$ using the above parametrization together with Eqs.~(\ref{eq: A generic}) and (\ref{eq: eta A}). Note that the coefficient $c$ is canceled out in $\eta^{\rm EXPP}_R$.  Also, the transversality condition has been imposed~\cite{Asner:2003gh}, and we find that relaxing the condition yields similar results.

The EXPP and the QCDF approaches generally have different shapes in the differential rates and, hence, render different values of $\eta_R$, {\it i.e.}, $\eta^{\rm EXPP}_R\neq\eta^{\rm QCDF}_R$.
Therefore, the EXPP of the normalized differential rates should be contrasted with the theoretical predictions ({\it e.g.}, from QCDF calculation) as the latter properly takes into account the energy dependence in weak decay amplitudes.

{\it Example ---}
There are many particles with mass around 1~GeV that $B$ mesons can decay into and have sufficiently large decay widths.  Here we consider an explicit example of the $B^- \to \rho(770)\pi^- \to \pi^-(p_1)\pi^+(p_2)\pi^-(p_3)$ decay to show the width effects associated with the $\rho$ meson, for which $\Gamma_\rho/m_\rho=0.192$.  In QCDF, its decay amplitude has the expression~\cite{Cheng:2020ipp}
\begin{align}
\label{eq:rhoamp}
\begin{split}
{\cal A}_{\rho}
=&
-g^{\rho\pi\pi} F(s_{12},m_\rho) \tilde A(B\to \rho(m_{12})\pi)
T_{\rho}^{\rm GS}(s_{12})
\\
& \quad \times 2q\cos\theta
+ (s_{12}\leftrightarrow s_{23})
~,
\end{split}
\end{align}
where $g^{\rho\pi\pi}$ is the strong coupling mediating the physical $\rho\to \pi^+\pi^-$ decay, $q=|\vec{p}_1|=|\vec{p}_2|$ is the momentum of either pion and $\theta$ is the angle between $\vec{p}_1$ and $\vec{p}_3$ in the rest frame of $\rho$, $s_{ij} \equiv (p_i + p_j)^2$ $(i,j = 1,2,3)$ and $T_\rho^{\rm GS}$ is the line shape of $\rho$ to be introduced below.  A form factor $F(s_{12},m_\rho)$ is introduced here to take into account the off-shell effect of $\rho(m_{12})\to\pi^+\pi^-$ when $m_{12}$ is off from $m_\rho$.  We shall follow Ref.~\cite{Cheng:FSI} to parameterize the form factor as $F(s,m_R)=[( {\Lambda^2+m_R^2)/( \Lambda^2+s})]^n$, where the cutoff $\Lambda$ is not far from the resonance, $\Lambda=m_R+\beta\Lambda_{\rm QCD}$ with $\beta = {\cal O}(1)$.  We shall use $n = 1$, $\Lambda_{\rm QCD}=250$~MeV and $\beta=1.0\pm0.2$ in subsequent calculations.

The quasi-two-body decay amplitude is given by \cite{CCC}
\begin{eqnarray}
\label{eq:Arhopi_bar}
&&
\tilde A(B^-\to\rho^0(m_{12})\pi^-)
\nonumber \\
&&=
\frac{G_F}{2}\sum_{p=u,c}\lambda_p^{(d)}\Bigg\{  \Big[\delta_{pu}(a_2-\beta_2)-a_4^p-r_\chi^\rho a_6^p
\nonumber \\
&&\quad
+{3\over 2}(a_7^p+a_9^p)+{1\over 2}(a_{10}^p+r_\chi^\rho a_8^p)
-\beta_3^p-\beta^p_{\rm 3,EW}\Big]_{\pi\rho}
\nonumber \\
&&\quad
\times \tilde X^{(B\pi,\rho)}
+  \Big[\delta_{pu}(a_1+\beta_2)+a_4^p-r_\chi^\pi a_6^p+a_{10}^p
\nonumber \\
&&\quad
-r_\chi^\pi a_8^p
+\beta_3^p+\beta^p_{\rm 3,EW}\Big]_{\rho\pi}\tilde X^{(B \rho,\pi)}\Bigg\}
~,
\end{eqnarray}
where $\lambda_p^{(d)}\equiv V_{pb} V^*_{pd}$ is the CKM factor.
The chiral factors $r_\chi^\rho$ and $r_\chi^\pi$ are defined in Ref.~\cite{BN}.
The factorizable matrix elements read
\begin{align}
\begin{split}
\tilde X^{(B\pi,\rho)}
&= 2f_\rho m_B \tilde p_c F_1^{B\pi}(s_{12})
~,
\\
\tilde{X}^{\left(B \rho, \pi\right)}
&= 2 f_{\pi} m_{B} \tilde{p}_{c} \Big[A_{0}^{B \rho}\left(m_{\pi}^{2}\right)
+ \frac{1}{2 m_{\rho}}
\\
&\quad
\times \left(m_{B}-m_{\rho}-\frac{m_{B}^{2}-s_{12}}{m_{B}+m_{\rho}}\right) A_{2}^{B \rho}\left(m_{\pi}^{2}\right)\Big]
~,
\end{split}
\end{align}
with $\tilde p_c = [(m_B^2-m_\pi^2-s_{12})^2 - 4m_B^2m_\pi^2]^{1/2}/(2m_B)$.

In Eq.~\eqref{eq:rhoamp}, the broad $\rho(770)$ resonance is commonly described by the Gounaris-Sakurai model~\cite{Gounaris:1968mw}, as employed by both BaBar \cite{BaBarpipipi} and LHCb \cite{Aaij:3pi_1,Aaij:3pi_2} in their analyses involving the $\rho(770)$ resonance
\begin{align}
\begin{split}
T_\rho^{\rm GS}(s)
&=
{ 1+D\Gamma_\rho^0/m_\rho \over s-m^2_{\rho}-f(s)+im_{\rho}\Gamma_{\rho}(s)}
~,
\end{split}
\label{eq:GS model}
\end{align}
where $\Gamma_\rho^0 \equiv \Gamma_\rho(m_\rho^2)$ and explicit expressions of $\Gamma_{\rho}(s)$, $D$, and $f(s)$ can be found in Refs.~\cite{BaBarpipipi,Aaij:3pi_1,Aaij:3pi_2}.  Note that $f(s)$ takes into account the real part of the pion-pion scattering amplitude with an intermediate $\rho$ exchange calculated from the dispersion relation.
In the NWA, $\Gamma_\rho\to 0$ and $s_{12}\to m_\rho^2$. It is straightforward to show that $\tilde A(B^-\to \rho(m_{12})\pi^-)$ is reduced to the QCDF amplitude of the physical $B^-\to \rho\pi^-$ process given in \cite{BN}. The decay rate is given by
\begin{eqnarray}
&&\Gamma(B^-\to \rho\pi^-\to \pi^+\pi^-\pi^-) \nonumber
\\
&&~~= {1\over 2}\,{1\over(2\pi)^3 32 \,m_B^3}\int ds_{23}\,ds_{12}|{\cal A}_\rho|^2. 
\end{eqnarray}
After integrating out the angular distribution, 
we are led to the desired factorization relation
\begin{eqnarray}
&& \Gamma(B^-\to \rho\pi^-\to \pi^+\pi^-\pi^-) \nonumber \\
&& \xlongrightarrow[]{\; \Gamma_{\rho}\to 0 \;}
 \Gamma(B^-\to \rho\pi^-) {\cal B}(\rho\to \pi^+\pi^-),
\end{eqnarray}
where use of the relations
\begin{eqnarray}
 \Gamma_{\rho\to \pi^+\pi^-} &=& {q_0^3\over 6\pi m_{\rho}^2}g_{\rho\to \pi^+\pi^-}^2, \nonumber\\
 \Gamma_{B^-\to \rho\pi^-} &=& {p_c\over 8\pi m_B^2}|A(B^-\to \rho\pi^-)|^2,
\end{eqnarray}
has been made.

%====================================================================
\begin{figure}[th]
\centering
\includegraphics[width=0.43\textwidth]{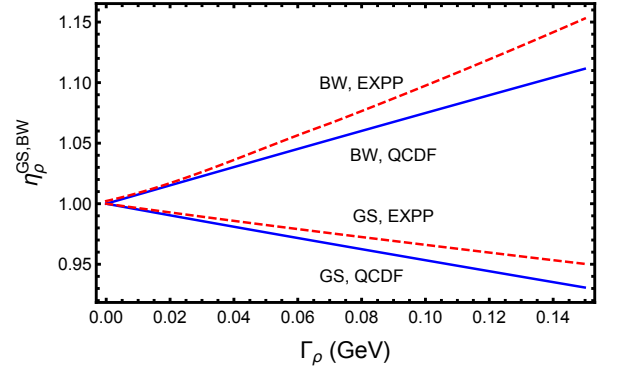}
\caption{$\eta_\rho$ as a function of the $\rho$ width in the Gounaris-Sakurai and the Breit-Wigner models, where the solid (dashed) curves come from the QCDF (EXPP) calculation.
}
\label{fig:eta_rho}
\end{figure}
%=====================================================================

For the finite width $\Gamma_\rho^0=149.1\pm0.8$~MeV, we find for the $B^-\to \rho\pi^-\to \pi^+\pi^-\pi^-$ decay that
\begin{align}
\label{eq:BFCPrho}
\begin{split}
{\cal B}
&= (8.76^{+1.86}_{-1.68})\times 10^{-6}
~,
~A_{C\!P}
= -(0.24^{+0.46}_{-0.54})\%
~,
\end{split}
\end{align}
and (with negligible errors)
\begin{align}
\eta_{\rho}^{\rm GS,QCDF}=0.931 ~~~(0.855)
~,
\end{align}
where the number in parentheses is obtained with the form factor $F(s,m_\rho) = 1$.
The deviation of $\eta_\rho^{\rm GS}$ from unity at 7\% level is contrasted with the ratio $\Gamma_\rho/m_\rho=0.192$. For comparison, using the Breit-Wigner model to describe the $\rho$ line shape would give
\begin{align}
\eta_{\rho}^{\rm BW,QCDF}=1.111\pm0.001 ~~~(1.033)
~.
\end{align}
In the EXPP scheme, on the other hand, we obtain
\begin{align}
\begin{split}
\eta_{\rho}^{\rm GS,EXPP} = 0.950\,,~~
\eta_{\rho}^{\rm BW,EXPP} = 1.152\pm0.001
~.
\end{split}
\end{align}
The dependence of $\eta_\rho$ as a function of the $\rho$ width, be it variable, is depicted in Fig.~\ref{fig:eta_rho} for both  Gounaris-Sakurai and Breit-Wigner line shape models.
It is the numerator of Eq.~\eqref{eq:GS model} that accounts for the result $\eta_\rho^{\rm GS}<1<\eta_\rho^{\rm BW}$. Since the Gounaris-Sakurai line shape was employed by both BaBar and LHCb in their analyses, ${\cal B}(B^-\to \rho\pi^-)$ should be corrected using  $\eta_\rho^{\rm GS}$ rather than $\eta_\rho^{\rm BW}$.

{\it Discussions ---}
In Table~\ref{tab:eta}, we give a summary of the $\eta_R$ parameters calculated in both QCDF and EXPP approaches for various resonances produced in some three-body $B$ decays.  Since the strong coupling mediating the $R(m_{12})\to P_1P_2$ decay is suppressed by $F(s_{12},m_R)$ when $m_{12}$ is off the $m_R$ shell, this implies a suppression in the three-body decay rate, rendering $\eta_R^{\rm QCDF}$ always larger than $\bar\eta_R^{\rm QCDF}$, which is defined for $F(s_{12},m_R) = 1$.  We see from Table~\ref{tab:eta} that this off-shell effect is small in vector mesons, but prominent in $K_2^*(1430)$, $\sigma/f_0(500)$ and $K_0^*(1430)$.  Moreover, the parameters $\eta_R^{\rm QCDF}$ and $\eta_R^{\rm EXPP}$ are similar for vector mesons, but different in the production of tensor and scalar resonances.

\begin{figure}[!th]
\centering
\includegraphics[width=0.43\textwidth]{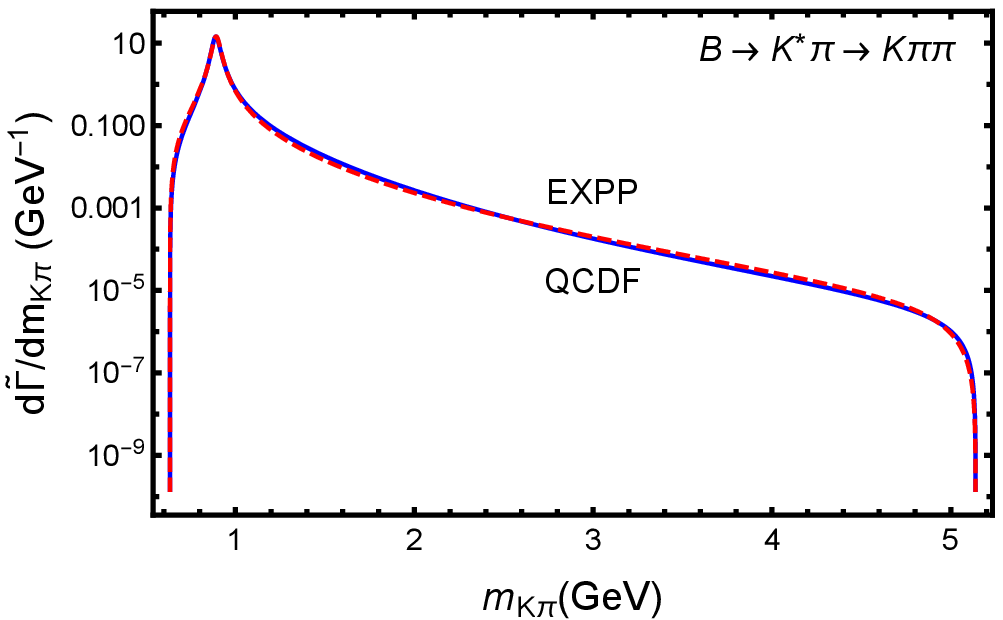}
\\
\vspace{4mm}
\includegraphics[width=0.43\textwidth]{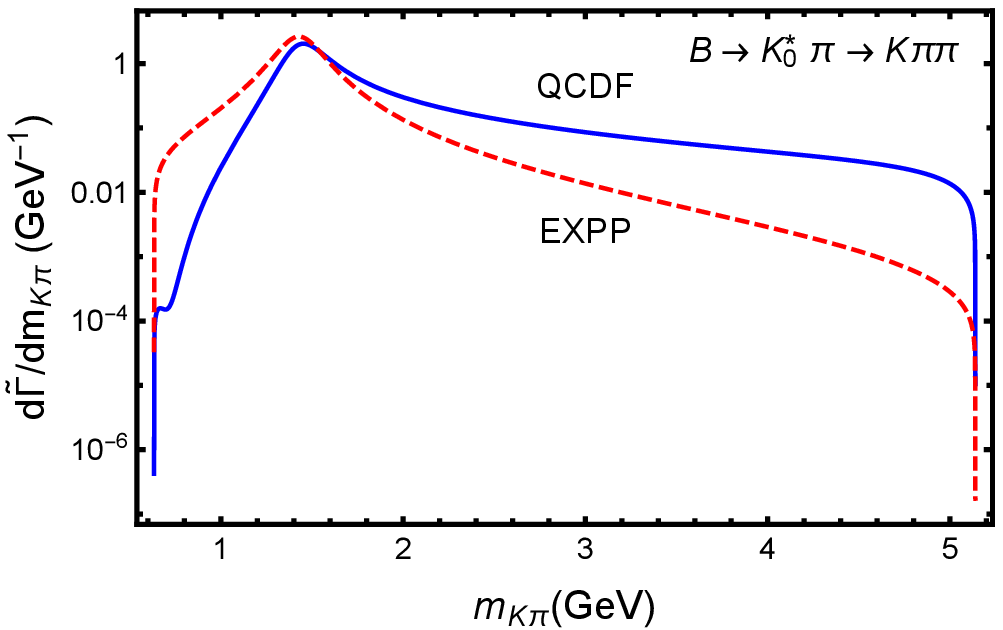}
\caption{Normalized differential rates (NDRs)
in $B^-\to  \bar K^{*0}\pi^-\to K^-\pi^+\pi^-$ and $B^-\to \bar K^{*0}_0\pi^-\to K^-\pi^+\pi^-$ decays.}
\label{fig: dGam}
\end{figure}

Take the $K^*(890)$ resonance as an example.  The NDRs in the QCDF calculation and the EXPP scheme are alike, as shown in the upper plot of Fig~\ref{fig: dGam}.  Consequently, the $\eta^{\rm QCDF}_{K^*}$ and $\eta^{\rm EXPP}_{K^*}$ are similar as expected from Eq.~\eqref{eq: eta A}.
While for $K^*_0(1430)$, as the values of NDR are anticorrelating, we can relate the smallness of $\eta^{\rm QCDF}_{K^*_0}$ relative to $\eta^{\rm EXPP}_{K^*_0}$ to the fact that the NDR obtained in the QCDF calculation is much larger than that using the EXPP scheme in the off-resonance region (particularly for large $m_{K\pi}$), as shown in the lower plot of Fig.~\ref{fig: dGam}.

%====================================================================
%=====================================================================

To understand the enhancement of the $B^-\to \overline K_0^{*}(1430)^0 \pi^-\to K^-\pi^+\pi^-$ NDR in the large $m_{K\pi}$ region in the QCDF calculation, we need to look into the following amplitude,
\begin{align}
\label{eq:K0stamp}
\begin{split}
&
{\cal A}_{K_0^*(1430)}
\\
&=
{G_F\over \sqrt{2}}\sum_{p=u,c}\lambda_p^{(s)}g^{K_0^*\to K^-\pi^+} F(m^2_{K\pi},m_{K_0^*})
\\
&\quad
\times T_{K_0^*}^{\rm BW}(m^2_{K\pi})
\Bigg[a_4^p-{1\over 2}a_{10}^p
+ \delta_{pu}\beta_2^p
+\beta_3^p
\\
&\qquad\quad
+\beta^p_{\rm 3,EW}
- r_\chi^{K_0^*}
\Big({m^2_{K\pi}\over m_{K_0^*}^2}\Big)
\Big(a_6^p-{1\over 2} a_8^p\Big) \Bigg]_{\pi K^*_0}
\\
&\quad
\times
f_{\bar K_0^*}
F_0^{B\pi}(m^2_{K\pi})
(m_B^2-m_\pi^2)
~,
\end{split}
\end{align}
where $g^{K_0^*\to K^-\pi^+} $ is the strong coupling constant for the physical $\overline K_0^{*}(1430)^0 \to K^-\pi^+$ decay,
$F(m^2_{K\pi},m_{K_0^*})$ is the strong decay form factor,
$T_{K_0^*}^{\rm BW}(m^2_{K\pi})$ is the BW factor,
$F_0^{B\pi}(m^2_{K\pi})$ is the $B\to \pi$ form factor,
$r^{K^*_0}_\chi$ is the chiral factor
and $f_{\bar K_0^*}$  is the vector decay constant of $\overline K_0^*(1430)$.
From the above equation, we can clearly see that the $m_{K\pi}$ dependence in the QCDF amplitude is governed by the strong decay form factor, $F(m^2_{K\pi},m_{K_0^*})$, the $B\to \pi$ form factor, $F_0^{B\pi}(m^2_{K\pi})$ and a factor of $m_{K\pi}^2/m_{K_0^*}^2$, in additional to the BW factor, $T_{K_0^*}^{\rm BW}(m^2_{K\pi})$.  The last two factors, namely $F_0^{B\pi}(m^2_{K\pi})$ and $m_{K\pi}^2$, are responsible for the enhancement of the QCDF differential rate in the large $m_{K\pi}$ region, and are not included in the EXPP approach for the scalar resonance.  As a result, QCDF and EXPP give different NDRs and $\eta_R$'s for this mode.

In the EXPP scheme, the weak dynamics is parameterized by a constant complex number, $c$, without any momentum dependence.  In the NWA, the normalized differential rate is highly peaked around the resonance and much suppressed elsewhere.  Therefore, it is justified to use a momentum-independent coefficient to represent the weak dynamics.  However, in the case of a broad resonance, things are generally different because the peak at the resonance is no longer dominating, with its height lowered by the NDR elsewhere.  In this case, the momentum dependence in the weak dynamics cannot be ignored and, hence, using a momentum-independent coefficient to represent the weak dynamics is too na\"ive.

As shown in Table~\ref{tab:eta}, the finite-width effects are significant in the $B^+\to \rho\pi^+$ decay and prominent in $B^+\to\sigma/f_0(500)\pi^+$ and $B^+\to K_0^{*0}(1430)\pi^+$.  For instance, the PDG values of~\cite{PDG}
\begin{align}
\begin{split}
{\cal B}(B^+ \to \rho \pi^+)^{\rm PDG} &= (8.3\pm1.2)\times 10^{-6}
~,
\\
{\cal B}(B^+ \to K^{*0}_0 \pi^+)^{\rm PDG} &= (39^{+6}_{-5}) \times 10^{-6}
\end{split}
\end{align}
should be corrected to
\begin{align}
\begin{split}
{\cal B}(B^+ \to \rho \pi^+)^{\rm EXPP} &= (7.9\pm1.1)\times 10^{-6}
~,
\\
{\cal B}(B^+ \to K^{*0}_0 \pi^+)^{\rm EXPP} &= (43^{+7}_{-6}) \times 10^{-6}
\end{split}
\end{align}
in the EXPP scheme and 
\begin{align}
\begin{split}
{\cal B}(B^+ \to \rho \pi^+)^{\rm QCDF} &= (7.7\pm1.1)\times 10^{-6}
~,
\\
{\cal B}(B^+ \to K^{*0}_0 \pi^+)^{\rm QCDF} &= (32^{+5}_{-4}) \times 10^{-6}
\end{split}
\end{align}
in the QCDF approach.

{\it Summary ---}
We have presented a general framework for computing the correction factor $\eta_R$ for properly extracting quasi-two-body decay rates from resonant three-body decay rates when the resonance has a sufficiently large width, and shown that it is given by the value of the normalized differential decay rate evaluated at the resonance.  The shape of the normalized differential rate thus matters in the determination of $\eta_R$.  We point out that the usual experimental parameterization ignores momentum dependence in weak dynamics and would lead to incorrect extraction of quasi-two-body decay rates in the case of broad resonances, as contrasted with the estimates using the QCDF approach.  Among the studied processes, the difference between the two approaches ranges from a few to $\sim 40\%$.

{\it Acknowledgments---}
This work was supported in part by the Ministry of Science and Technology (MOST) of Taiwan under Grant Nos. MOST-108-2112-M-002-005-MY3 and MOST-106-2112-M-033-004-MY3.

%====================================================================
\begin{widetext}
\begin{table*}[t]
\caption{A summary of the $\eta_R$ parameter for various resonances produced in the three-body $B$ decays. Off-shell effects on the strong coupling $g^{RP_1P_2}$ are taken into account in the determination of $\eta_R^{\rm QCDF}$ but not in $\bar\eta_R^{\rm QCDF}$.  Uncertainties in $\eta_R$ are not specified whenever they are negligible.}
\vskip 0.07cm
\label{tab:eta}
\footnotesize{
\begin{ruledtabular}
\begin{tabular}{ l l c c l l l}
 Resonance~~~ & ~$B^+\to RP_3\to P_1P_2P_3$ ~~~ & ~$\Gamma_R$ (MeV)~\cite{PDG}~ & $\Gamma_R/m_R$ & ~~~$\bar\eta_R^{\rm QCDF}$ & ~~~$\eta_R^{\rm QCDF}$ & ~~~$\eta^{\rm EXPP}_R$ \\
\hline
$f_2(1270)$ & $B^+\to f_2\pi^+\to \pi^+\pi^-\pi^+$ & ~$186.7^{+2.2}_{-2.5}$~~ & 0.146 & ~~0.974 &  ~~$1.003^{+0.001}_{-0.002}$ & ~~$0.937^{+0.006}_{-0.005}$ \\
$K_2^*(1430)$ & $B^+\to K^{*0}_2\pi^+\to K^+\pi^-\pi^+$ & ~$109\pm5$~~ & 0.076 & ~~$0.715\pm0.009$ & ~~$0.972\pm0.001$ & ~~$1.053\pm0.002$ \\
$\rho(770)$ & $B^+\to \rho^0\pi^+\to \pi^+\pi^-\pi^+$ & ~$149.1\pm0.8$~~ & 0.192 & ~~0.86 (GS) & ~~0.93 (GS) & ~~0.95 (GS)\\
 &  &  & & ~~1.03 (BW) & ~~1.11 (BW) & ~~1.15 (BW)\\
$\rho(770)$ & $B^+\to K^+\rho^0 \to K^+\pi^+\pi^-$ & ~$149.1\pm0.8$~~ & 0.192 & ~~0.90 (GS) & ~~0.95 (GS) & ~~0.93 (GS)\\
 &  &  & & ~~1.09 (BW) & ~~1.13 (BW) & ~~1.13 (BW)\\
$K^*(892)$ & $B^+\to K^{*0}\pi^+\to K^+\pi^-\pi^+$ & ~$47.3\pm0.5$~~ & 0.053 & ~~1.01 & ~~$1.067\pm0.002$ & ~~1.075 \\
$\sigma/f_0(500)$ & $B^+\to \sigma\pi^+\to \pi^+\pi^-\pi^+$ & ~~~~$700\pm26$ \cite{Aaij:3pi_2} & $\approx 1.24$ & ~~$1.63\pm0.03$ & ~~$2.15\pm0.05$ & ~~$1.50\pm0.02$     \\
$K_0^*(1430)$ & $B^+\to K^{*0}_0\pi^+\to K^+\pi^-\pi^+$ & ~$270\pm80$~~ & $\approx 0.189$ &~~$0.31^{+0.08}_{-0.05}$ & ~~$0.83\pm0.04$ & ~~$1.11\pm0.03$ \\
\end{tabular}
\end{ruledtabular} }
\end{table*}
\end{widetext}
%====================================================================

%\vspace{-3mm}


\begin{thebibliography}{}

\bibitem{PDG}
P. A. Zyla {\it et al.} [Particle Data Group], Prog. Theor. Exp. Phys. 2020, 083C01 (2020).

\bibitem{BBNS} M. Beneke, G. Buchalla, M. Neubert, and C.T. Sachrajda,
%``QCD factorization for $B \to PP$ decays: Strong phases and CP violation in the heavy quark limit,''
Phys. Rev. Lett. \textbf{83}, 1914 (1999);
%doi:10.1103/PhysRevLett.83.1914
%[arXiv:hep-ph/9905312 [hep-ph]];
%``QCD factorization for exclusive, nonleptonic B meson decays: General arguments and the case of heavy light final states,''
Nucl. Phys. B \textbf{591}, 313 (2000).
%doi:10.1016/S0550-3213(00)00559-9
%[arXiv:hep-ph/0006124 [hep-ph]].

\bibitem{CCC}
H.~Y.~Cheng, C.~W.~Chiang and C.~K.~Chua,
%``Finite-Width Effects in Three-Body B Decays,''
arXiv:2011.07468 [hep-ph].

\bibitem{Asner:2003gh}
D.~Asner,
%``Charm Dalitz plot analysis formalism and results: Expanded RPP-2004 version,''
arXiv:hep-ex/0410014 [hep-ex].

\bibitem{Gounaris:1968mw}
G.~J.~Gounaris and J.~J.~Sakurai,
%``Finite width corrections to the vector meson dominance prediction for $\rho \to e^+e^-$,''
Phys. Rev. Lett. \textbf{21}, 244 (1968).
%doi:10.1103/PhysRevLett.21.244

\bibitem{Cheng:2020ipp}
H.~Y.~Cheng and C.~K.~Chua,
%``Branching fractions and $CP$ violation in $B^-\to K^+K^-\pi^-$ and $B^-\to \pi^+\pi^-\pi^-$ decays,''
Phys. Rev. D \textbf{102},  053006 (2020).
%doi:10.1103/PhysRevD.102.053006
%[arXiv:2007.02558 [hep-ph]].

\bibitem{Cheng:FSI}
H.~Y.~Cheng, C.~K.~Chua and A.~Soni,
%``Final state interactions in hadronic $B$ decays,''
Phys. Rev. D \textbf{71}, 014030 (2005).
%doi:10.1103/PhysRevD.71.014030
%[arXiv:hep-ph/0409317 [hep-ph]].

\bibitem{BN} M. Beneke and M. Neubert,
%``QCD factorization for $B\to PP$ and $B \to PV$ decays,''
Nucl. Phys. B \textbf{675}, 333 (2003).
%doi:10.1016/j.nuclphysb.2003.09.026
%[arXiv:hep-ph/0308039 [hep-ph]].

\bibitem{BaBarpipipi}
  B.~Aubert {\it et al.}   [BaBar Collaboration],
%  ``Dalitz Plot Analysis of $B^+\to \pi^+\pi^+\pi^-$ Decays,''
  Phys.\ Rev.\ D {\bf 79}, 072006 (2009).%  [arXiv:0902.2051 [hep-ex]].

\bibitem{Aaij:3pi_1}
  R.~Aaij {\it et al.} [LHCb Collaboration],
%  ``Observation of Several Sources of $CP$ Violation in $B^+ \to \pi^+ \pi^+ \pi^-$ Decays,''
  Phys.\ Rev.\ Lett.\  {\bf 124},  031801 (2020).
  %doi:10.1103/PhysRevLett.124.031801
%  [arXiv:1909.05211 [hep-ex]].

\bibitem{Aaij:3pi_2}
  R.~Aaij {\it et al.} [LHCb Collaboration],
%  ``Amplitude analysis of the $B^+ \rightarrow \pi^+\pi^+\pi^-$ decay,''
  Phys.\ Rev.\ D {\bf 101},  012006 (2020).
  %doi:10.1103/PhysRevD.101.012006
%  [arXiv:1909.05212 [hep-ex]].



\end{thebibliography}
\end{document}